# Sub-20 kHz low frequency noise near ultraviolet miniature external cavity laser diode

R. Kervazo[1], A. Congar[2], G. Perin[1], L. Lablonde[3], R. Butté[4], N. Grandjean[4], L. Bodiou[1], J. Charrier[1], S. Trebaol[1]

[1] *Univ Rennes, CNRS, Institut FOTON - UMR 6082, F-22305 Lannion, France*

[2] *Oxxius, 4 rue Louis de Broglie, 22300 Lannion, France*

[3] *Exail, rue Paul Sabatier, Lannion, France*

[4] *Institute of Physics, École Polytechnique Fédérale de Lausanne (EPFL), CH-1015 Lausanne, Switzerland*

*We present a compact InGaN fiber Bragg grating (FBG) semiconductor laser diode operating below 400 nm in the single-mode emission regime. This compact coherent laser source exhibits an intrinsic linewidth of 14 kHz in the near-UV range and a side-mode suppression ratio reaching up to 40 dB accompanied by a mW-level output power. Furthermore, the properties of the FBG, including its central wavelength, bandwidth, and reflectivity, can be readily customized to fulfill specific requirements. As a result, the small footprint design of this laser is compatible with integration into a standard butterfly package to ease the lab-to-market technology transfer. The combination of low frequency noise and fibered output signal positions these FBG laser systems as strong candidates for hybridization with integrated photonic platforms tailored for quantum information processing and metrology.*

Nowadays, laser diodes are essential elements for the development of compact photonic devices. They offer the opportunity of combining good performances in terms of linewidth, optical power and frequency noise [1–3]. The needs in the field of optical telecommunications have enabled the development of compact laser diodes mainly in the telecom C-band. The implementation of such compact laser diodes with low-frequency noise at visible wavelengths is a technological challenge recently addressed by several research groups [4–8]. The objective is to provide laser diodes capable of meeting several application requirements in the fields of classical and quantum sensors and the metrology of optical frequencies for the production of compact optical clocks [9–11]. Indeed, such clocks require the use of narrow-spectrum and low-frequency noise lasers at visible and near-ultraviolet wavelengths for cooling, capturing and manipulating atoms or ions [10,12–14]. Commercial laser devices offering the necessary performances are usually based on an external cavity involving a diffraction grating that does not provide the desired compactness and mechanical robustness for applications outside laboratories. However, recent demonstrations based on compact external cavities, such as integrated optical resonators [4,5] offer interesting perspectives. Despite meeting the compactness requirement, these approaches suffer from reduced figures of merit both in terms of output power (in the µW range) and linewidth (in the MHz range). Moreover, the performance of state-of-the-art optical clocks is still limited by the frequency noise floor of the lasers used [15].
Indeed, efforts are needed to lower this noise floor by increasing the finesse of the external cavity, which can be obtained by reducing the propagation losses and by increasing the reflectivity of the output mirror. Fiber Bragg gratings [16] (FBGs) are suitable devices for deporting the external cavity in the collection fiber, combining low propagation losses and a wavelength-selective mirror whose reflectivity and bandwidth can be widely adjusted to find the right optical power/finesse trade-off. Moreover, the Bragg center wavelength can be finely adjusted by less than 0.1 nm with current fabrication methods [17]. In addition, fiber gratings exhibit a significantly lower sensitivity to temperature fluctuations compared to laser diode waveguides [18]. This characteristic provides the opportunity to operate without thermal regulation. Initially implemented in the C-band for wavelength-division multiplexing applications [19,20], research works on fiber Bragg grating lasers (FGLs) are now reported at various optical wavelengths [7,21]. Indeed, due to their exceptional wavelength accuracy and stability, fiber gratings can serve as frequency references, enabling single-mode laser frequency emissions that precisely align with the requirements of atomic spectroscopy and metrology [9,22].

Despite interesting performances in terms of output power and low frequency noise, previously reported FGL configuration in the near-UV [7] suffered from a lack of integration due to the use of lenses for beam shaping. Indeed, looking for the optimum trade-off between footprint reduction, ease of integration from one side, and optical stability, output power optimization from the other side speak in favor of a butt-coupled configuration of the FGL. Historically, such an optimized laser configuration has been first implemented in the telecom band, which allowed the lab-to-market technology transfer known as the compact butterfly packaging configuration that became the standard for large scale photonic integration.

In the present work, we propose to extend this approach to the near-UV spectral range by demonstrating a miniature footprint configuration of a UV laser diode emitting at 399.6 nm subject to optical feedback from an FBG. The resulting 2 cm external fibered cavity acts as a spectral filter that strongly improves the spectral performance of the initially multimode laser (Fig. 1). Even though shortening the operating wavelength implies an increased complexity in terms of optical alignment, the resulting single-mode laser displays an intrinsic linewidth down to 14 kHz with a mW-fibered-output power together with a 40 dB side-mode suppression ratio (SMSR) thanks to the use of a narrowband FBG. The structure of the laser diode, the fabrication and characterization steps of the Bragg grating contributing to the external cavity are first described. Then, the spectral performances of the laser, in particular the spectrum and frequency noise measurements, are presented.

The architecture of the single-mode laser diode (LD) is detailed on Fig. 1.

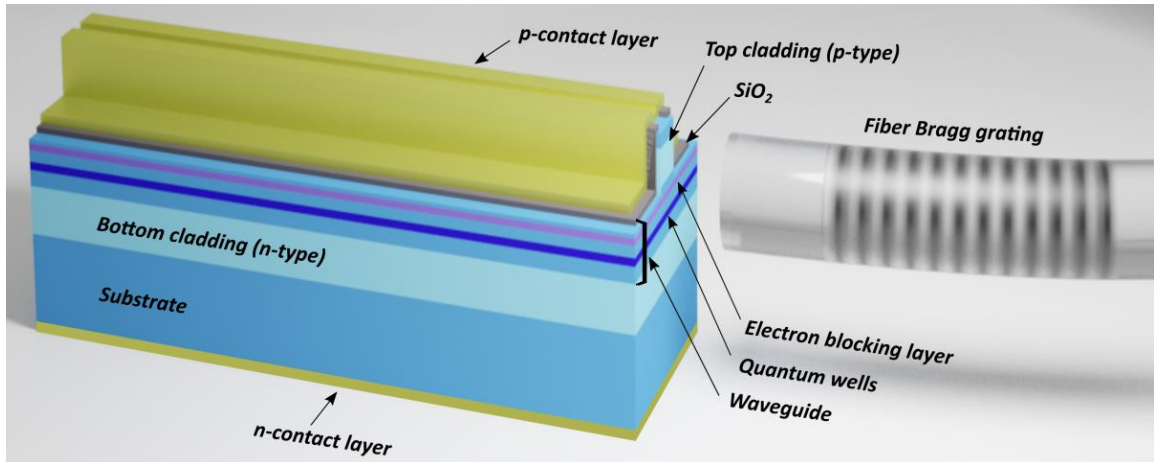

Fig. 1. Schematic of the FBG InGaN laser diode. Details are given in the Supplementary Material (Sec.1).

The laser diode stacking was epitaxially grown using the metalorganic vapor phase epitaxy method on a low threading dislocation density (~$10^6$ cm$^{-2}$) c-plane free-standing GaN substrate following the same procedure as described in ref. [23]. The optical gain, centered at 400 nm, is provided by two 4.5-nm-wide InGaN quantum wells separated by 12-nm-undoped In$_{0.05}$Ga$_{0.95}$N barriers, while a *p*-AlGaN electron-blocking layer is present on the p-type side of the junction to avoid any electron overflow. The bottom and top claddings are 1 µm and 500 nm thick, respectively. The epitaxial structure of the diode is exposed in the Supplementary Material (Sec.1). The laser consists of an 800-µm-long ridge structure, which corresponds to a measured Fabry-Perot cavity free spectral range (FSR) of 26 pm. The ridge width is limited to 2.5 µm to ensure single-mode transverse operation. A five-layer dielectric distributed Bragg reflector (SiO$_2$/Ta$_2$O$_5$) with a reflectivity $R_1$ of ~95% was deposited on the back mirror facet to maximize the emission directivity. On the opposite output facet, a single SiO$_2$ layer drops the reflection $R_2$ down to ~9% to improve the sensitivity to optical feedback from the FBG and consequently ease the single-mode operation of the laser. The laser diode temperature fluctuations are reduced to less than 0.01 °C using a Peltier controller. This temperature control allows the gain curve to be spectrally tuned by 0.07 nm/K on par with the gain frequency drift values reported in the literature for GaN laser diodes [24].

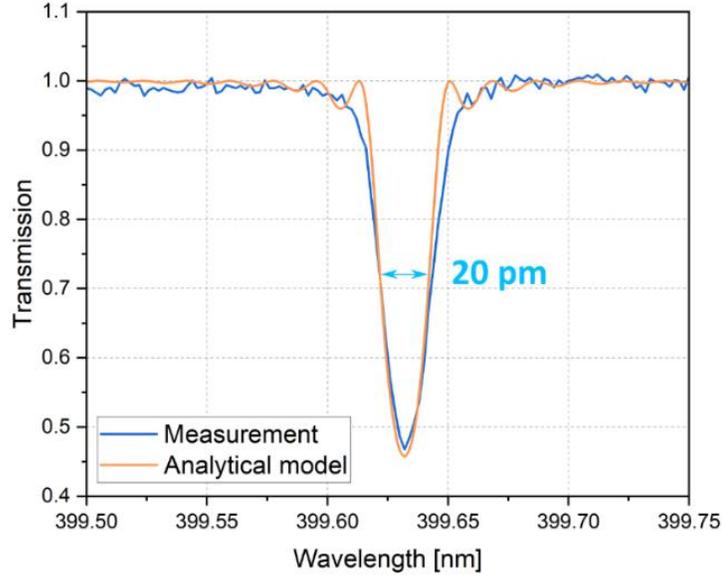

Fig. 2. Fiber Bragg grating transmission measurement (blue curve) and corresponding simulated spectrum (orange curve). The simulation parameters are the following ones: FBG length = 3 mm, index modulation $\Delta n = 0.40 \times 10^{-4}$. More details are given in the Supplementary Material (Sec.2).

An FBG is associated with the laser diode to form an external cavity. Thus, to obtain a stable single-mode operation and a spectral narrowing of the laser diode under optical feedback, only one longitudinal mode of the laser diode has to be selected by a full-width-at-half maximum (FWHM) of the Bragg grating reflectivity narrower than the FSR of the multimode laser diode. The FBG has been designed accordingly.

This mirror is fabricated by exploiting the photosensitivity of a short-wavelength commercially available germanosilicate single-mode fiber. The fiber is transversally exposed to a fringe pattern at a wavelength $\lambda_{uv} = 213$ nm using a 150-mW high-repetition rate nanosecond quintupled Nd:YVO$_4$ laser [25]. To address the issues for high-quality FBG over Bragg wavelengths ranging from 375 to 410 nm and to achieve Bragg wavelength flexibility, a scanning Talbot interferometer was used [26]. Usually, a direct phase mask writing technique is used to generate these fringes, but there is no accessible commercial solution yet to achieve a 400-nm Bragg wavelength. In a near future, femtosecond laser-written FBGs at short wavelengths could be investigated [27,28].

The transmission spectrum of the 3-mm FBG (Fig. 2, blue curve) displays a reflectivity $R_3$ of 53.5 % and a FWHM evaluated to 20 pm, close to the optical spectrum analyzer (OSA) resolution (10 pm). An analytical model based on coupled mode theory [16] is used to confirm the adequacy of the design with the measured transmission, as shown in Fig. 2 (orange curve). Further details on the simulation procedure are provided in the Supplementary Material (Sec.2).

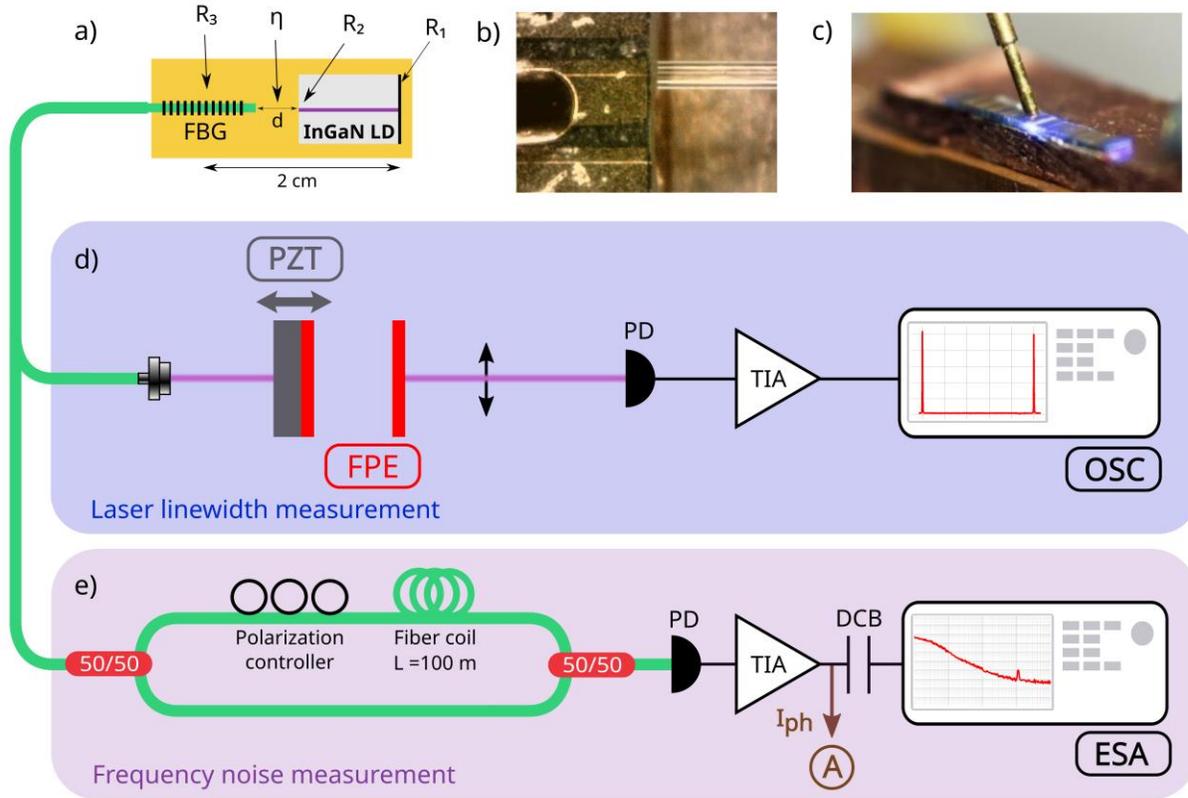

Fig. 3. UV FGL synopsis and characterization benches. a) UV FGL structure. The device is composed of an InGaN integrated laser diode with a back-facet reflectivity $R_1$ of ~95% and an output-facet reflectivity $R_2$ of ~9%. An FBG presenting a reflectivity $R_3$ of 53.5% is butt-coupled to the laser diode at a distance $d$ ~10 microns with a coupling coefficient $\eta$ of 23%. The total length of the cavity is less than 2 centimeters. b) Top view picture of the InGaN Fabry-Perot laser diode butt-coupled to the FBG. c) Picture of the UV Fabry-Perot laser under electrical pumping. d) Experimental setup for linewidth measurement based on a Fabry-Perot etalon (FPE). One of the mirrors is linearly translated, thanks to a piezo actuator (PZT), to scan the laser line. e) Frequency noise measurements are performed using a correlated delayed self-homodyne measurement setup composed of an imbalanced fiber-based Mach-Zehnder interferometer that plays the role of frequency noise discriminator. The electrical signals generated by photodiodes (PD) are directed to both an oscilloscope (OSC) and an electrical spectrum analyzer (ESA) via transimpedance amplifiers (TIA) and DC-blocks (DCB). For the measurement of DC photocurrents, a multimeter (A) is used.

The external FBG reflector is butt-coupled to the laser diode as shown in Fig. 3a). The length of the external cavity is less than 2 cm long and the residual gap $d$ between the cleaved fiber and the laser diode mirror output is evaluated to be less than 10 µm. The total footprint of the device corresponds to an area smaller than 0.5*2 cm² which is compatible with a commercial butterfly packaging. The optical feedback at the Bragg wavelength of the narrowband FBG aims to select the nearest longitudinal mode of the Fabry-Perot type semiconductor laser. This can be obtained since the FBG linewidth (20 pm) is narrower than the laser diode FSR (26 pm). Such butt-coupled approach avoids the use of bulky lenses as in extended external cavity approaches hence improving the robustness to mechanical vibrations and thus the spectral performances. The coupling coefficient in this configuration has been evaluated to $\eta$~23%. This value can be explained by the mismatch between the diode and the fiber mode profiles which do not present the same shapes (elliptical and circular, respectively). The experimental approach to determine the coupling efficiency and its impact on the laser performances are detailed in the Supplementary Material (Sec.3-5).

The optical spectrum of the laser diode was first characterized without Bragg grating (gray curve in Fig. 4a). Without feedback, the optical spectrum of the diode reveals a multimode operation with a maximum SMSR on the order of 20 dB. By carefully aligning the FBG in front of the Fabry-Perot laser diode output, the optical spectrum collapses in a stable single-mode operation. Thus, in its FGL configuration, the laser spectrum bandwidth is drastically reduced and exhibits a single-mode lasing operation at the Bragg wavelength with a 40 dB SMSR for a pump current of 124 mA. The FGL output power can reach up to 1.8 mW at 399.6 nm corresponding approximately to 3.9 mW at the InGaN LD output facet (see Supplementary Material Sec.6).

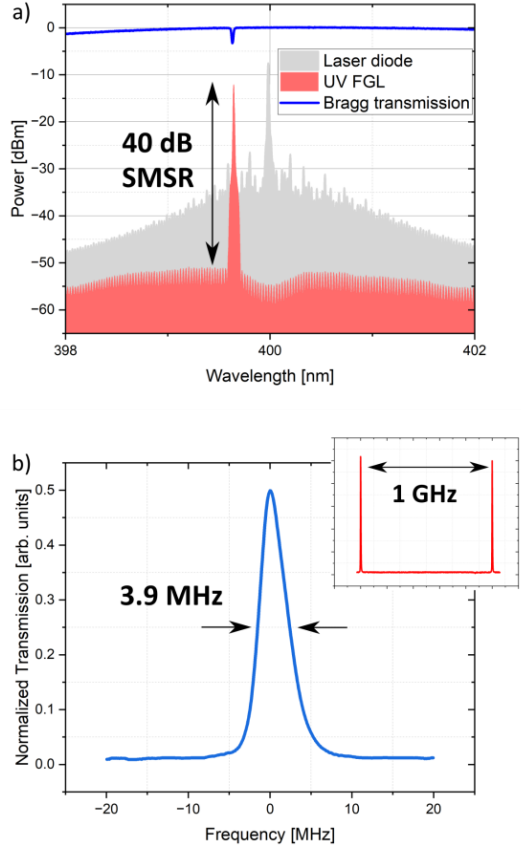

Fig. 4. Spectral characterization of the UV FGL. a) Optical spectra of the laser diode (gray curve) and the UV FGL (red curve) for a pump current of 124 mA. The blue curve corresponds to the fiber Bragg transmission centered at 399.6 nm. b) Optical spectrum of the UV FGL obtained with a FPE (3.9 MHz resolution and 1 GHz FSR). Inset) Optical spectrum measurement through the FPE confirming that the UV FGL behaves in the single longitudinal-mode regime.

Since the OSA resolution is not sufficient to prove single-mode operation, the spectral performances of the FGL are then studied using the characterization bench displayed in Fig. 3d). The high-resolution measurements of the optical spectrum were carried out with an Fabry-Perot etalon (FPE). The result is presented in the inset of Fig. 4b). Two peaks separated by 1 GHz, corresponding to the FSR of the FPE, confirmed that the FGL oscillates on a single mode of the external cavity. However, as the width of the measured peak (3.9 MHz) corresponds to the resolution of the FPE, this feature cannot be used to evaluate correctly the optical linewidth.

A self-homodyne [16] characterization bench based on a Mach-Zehnder [29] interferometer was built (Fig. 3e)) to determine the actual frequency noise (FN) of the laser. The description of the bench is given in supplementary material (Sec.7). This setup enables laser FN analysis over a frequency bandwidth ranging from 100 Hz to 2 MHz. The noise floor reaches up to 1000 $Hz^2/Hz$ at low frequencies and drop down to few $Hz^2/Hz$ above 100 kHz. Fig. 5a) shows the result of the measurements. The red curve represents the characterization of the FN of the UV FGL laser at 399.6 nm. For the sake of comparison, the FN of a commercial external cavity diode laser (ECDL, model Toptica DL Pro) used for Sr-ion cooling and Rb Rydberg excitation at 420.0 nm is shown (blue curve).

Usually, the frequency noise of a laser is characterized by a white noise of amplitude $h_0$ at high frequencies, which refers to the intrinsic linewidth of the laser and the noise at low Fourier frequencies whose slope changes in $\frac{1}{f^\alpha}$ with $1 < \alpha < 2$ corresponds to slow frequency drifts of the intrinsic linewidth [30]. This last contribution is associated with the Gaussian profile which, convolved with the Lorentzian shape intrinsic linewidth, gives the Voigt profile of the laser [31]. It is possible, from measurements of the FN, to estimate the laser linewidth using the Elliott formula [32] (see Supplementary Material Sec.8).

The FN of the FGL exhibits a plateau for frequency values above 300 kHz corresponding to the white noise associated with the intrinsic linewidth of the laser, which is mainly determined by the noise due to spontaneous emission. A value of $h_0 = 4149$ Hz$^2$/Hz gives an estimate of the intrinsic linewidth, through the Elliott formula, of the order of $14 \pm 2$ kHz. The corresponding reconstructed Lorentzian-shape intrinsic linewidth is shown in Fig. 5b) (red curve). The intrinsic linewidth is about one order of magnitude smaller than that of the Toptica DL Pro linewidth ($120 \pm 19$ kHz). The integration of the FN spectral density and in particular the low-frequency contribution gives access to an estimate of the Voigt profile with a linewidth on the order of $720 \pm 120$ kHz similar to the performances of the Toptica DL Pro. Indeed, acoustic and mechanical noises perturb the FGL central frequency, contributing to the broadening of the linewidth for longer integration times. Locking the laser to a more stable frequency reference [29] could provide a solution to overcome this technical noise affecting the emission frequency.

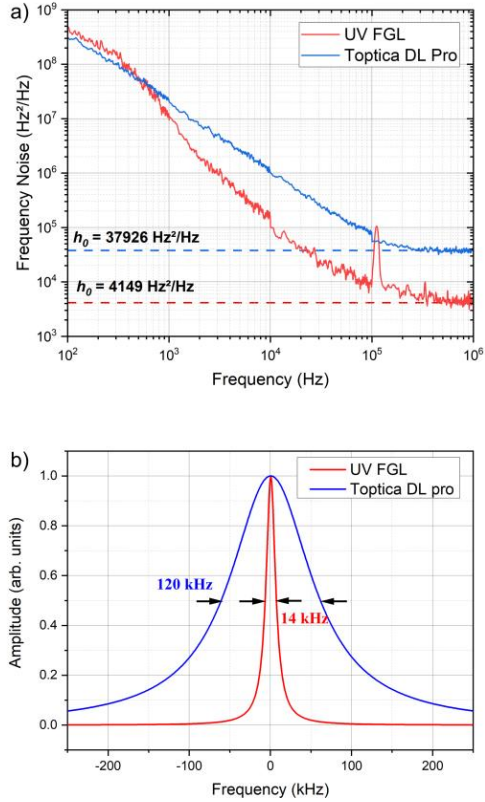

Fig. 5. a) Frequency noise measurements of the UV FGL at 399.6 nm (red color) and Toptica DL Pro emitting at 420.0 nm (blue color). The peak seen at ~100 kHz on the red curve is due to the photodiode noise. b) Reconstructed optical spectrum from frequency noise measurements by making use of the Elliott formula. Details are given in the Supplementary Material (Sec.8).

A comparison with similar hybrid or integrated ECDL systems is proposed in Table I. This comparison is based on a table-top commercial solution and state-of-the-art integrated ECDL lasers. The Toptica DL Pro is a renowned commercial ECDL proposing a high fibered output power of 45 mW with a narrow integrated linewidth of 887 kHz at 10 ms of integration time. Nevertheless, its dimensions prevent it from further integration in new commercial devices requiring high level of compactness.

Our purpose is to propose a more compact solution in the near-ultraviolet wavelength range. We have previously demonstrated an ECDL FGL laser [7] displaying a fibered output and good performances in terms of linewidth (950 kHz at 10 ms) and sufficient output power (1.3 mW) for the targeted applications. However, this configuration suffers from the use of discrete optical elements which prevents its compatibility with an integration in a commercial butterfly package. The footprint reduction of the device motivated the development of the new configuration reported in this paper.

To ensure compatibility with a butterfly package, the dimensions of the reported FGL laser were reduced by removing the lenses used for beam shaping. The presented device enables a significant improvement in terms of intrinsic linewidth surpassing previously reported demonstrations [5,6] based on compact external cavity semiconductor lasers by at least two orders of magnitude. This parameter is of primary importance to improve the locking performance of compact atomic clocks [15]. Moreover, the FGL laser output power can reach up to 1.8 mW at 399.6 nm, which is up to 100 times larger than the optical powers reported so far for such compact approaches in the near-UV range, which can be accounted for by the fact that integrated optics approaches [5,6] in this spectral range have suffered from large optical propagation losses that degrade the lasing performances.

Table I : Comparison of external cavity laser diodes emitting in the 400-420 nm range reported in the literature.

| Device / Research team | Year | ECDL Technology | $\lambda$ (nm) | SMSR (dB) | Fibered output power (mW) | Linewidth intrinsic | Linewidth integrated | Butterfly package compatibility |
|---|---|---|---|---|---|---|---|---|
| Toptica DL Pro | | Bulk Bragg grating | 420 | 47 | 45 | 150 kHz | 887 kHz at 10 ms | X |
| Congar et al. [7] | 2021 | Fiber Bragg grating | 400 | 44 | 1.3 | 16 kHz | 950 kHz at 10 ms | X |
| This work | 2024 | Fiber Bragg grating | 400 | 40 | 1.8 | **14 kHz** | **704 kHz at 10 ms** | ✓ |
| Corato-Zanarella et al. [5] | 2022 | Integrated microring | 404 | 21 | 0.02 | < 1.5 MHz (estimated) | 3.3 MHz | ✓ |
| Franken et al. [6] | 2023 | Integrated microring | 405 | 43 | 0.74 | | < 25 MHz | ✓ |

In summary, by butt-coupling an InGaN semiconductor laser diode emitting near 400 nm to an FBG, we demonstrated a compact narrow-linewidth semiconductor laser with a Lorentzian linewidth as low as 14 kHz corresponding to a white frequency noise below 5000 Hz$^2$/Hz. This mW-output power laser fulfills the low-frequency noise performances required for atomic spectroscopy and frequency metrology, offering the possibility to develop compact optical atomic clocks [10] and portable quantum sensors [33] based on electronic transitions occurring in the near-UV range [14] as for example, the excitation of the ytterbium Rydberg state at 399 nm for quantum information processing. The compactness, wavelength versatility and ease of access of the fibered output signal represent a promising trade-off between compactness and performances that should favor the hybridization of such an FGL with integrated photonic platforms for quantum information processing [34,35].

**Funding**

The present work is supported under projects UV4Life (contract N° 19005486), the European Regional Development Fund (contract N° EU000998), ANR COMBO (contract N° ANR-18-CE24-0003), Région Bretagne and Lannion Trégor Communauté.